\documentclass[twocolumn,showpacs,preprintnumbers]{revtex4}
\usepackage{amssymb}
\usepackage{graphicx}
\usepackage{dcolumn}
\usepackage{bm}

\begin{document}

\title{Probing the intrinsic state of a one-dimensional quantum well with a photon-assisted tunneling}
\author{S.~E.~Shafranjuk}
\homepage{http://kyiv.phys.northwestern.edu}
\affiliation{Department of Physics and Astronomy, Northwestern University, Evanston,
Illinois 60208.}
\date{\today }

\begin{abstract}
The photon-assisted tunneling (PAT) through a single wall carbon nanotube quantum well (QW) under influence an external electromagnetic field for probing of the Tomonaga Luttinger liquid (TLL) state is suggested. The elementary TLL excitations inside the quantum well are density ($\rho_{\pm}$) and spin ($\sigma_{\pm} $) bosons. The bosons populate the quantized energy levels $\varepsilon^{\rho + }_n =\Delta n/ g$ and $\varepsilon^{\rho -(\sigma \pm) }_n = \Delta n$ where $\Delta = h v_F /L $ is the interlevel spacing, $n$ is an integer number, $L$ is the tube length, $g$ is the TLL parameter. Since the electromagnetic field acts on the $\rho_{+}$ bosons only while the neutral $\rho_{-}$ and $\sigma_{\pm} $ bosons remain unaffected, the PAT spectroscopy is able of identifying the $\rho_{+}$ levels in the QW setup. The spin $\varepsilon_n^{\sigma+} $ boson levels in the same QW are recognized from Zeeman splitting when applying a d.c. magnetic field $H \neq 0$ field. Basic TLL parameters are readily extracted from the differential conductivity curves.
\end{abstract}

\pacs{73.23.Hk, 73.63.Kv, 73.40.Gk}
\maketitle

\section{Introduction}
\begin{figure}
 \includegraphics{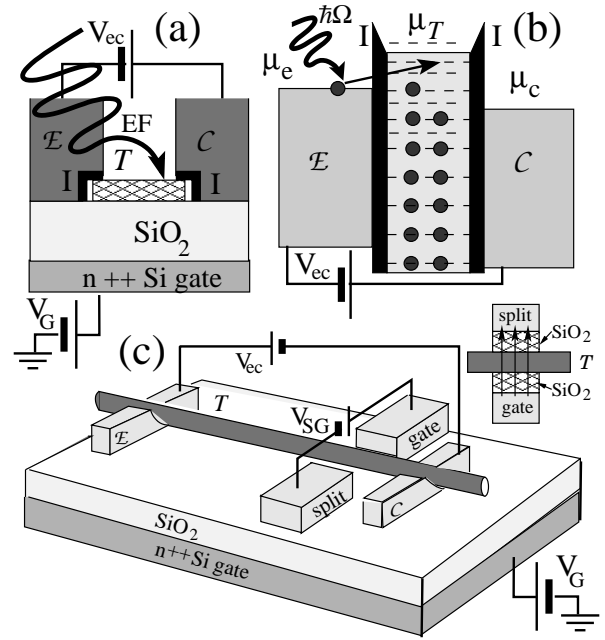}
 \caption{\label{fig:Setup_b}
(a) A quantum well (QW) composed of the 1D section $\mathcal{T}$ with attached emitter ($\mathcal E$) and
collector ($\mathcal C$) electrodes. The potential barriers are shown in
black at the $\mathcal{E}$/$\mathcal{T}$ and $\mathcal{T}$/$\mathcal{C}$
interfaces. (b) Energy diagram of the PAT process in the QW.
(c) The split gate configuration of the QW. The right side inset shows how
the electric field is applied to the $\mathcal{T}$ section.}
\end{figure}

A one dimensional quantum well (QW) gives many promises for scientific
research and various practical applications \cite%
{Dressel,chem,med,My-PRB,Paola-A,Paola-B}. One spectacular example is the
junction formed by a single wall carbon nanotube (SWCNT) with metallic
electrodes attached to its ends (see the sketch in Fig.~\ref{fig:Setup_b}). That setup
(see Fig.~\ref{fig:Setup_b}) harbors various condensed matter systems. Remarkable
properties\cite{Dressel,chem,med,My-PRB,Paola-A,Paola-B,Lambert,Babic,Cao,Wiel} of SWCNT
emerge from their intrinsic structure \cite{Dressel}. The single wall carbon $[n,m]$ nanotube is a rolled up atomic honeycomb monolayer formed by two sublattices A and B. The integer indices $n$ and $m$ ($n \geq m \geq 0$) of the rollup vector ${\bf R}=n {\bf R}_1 + m {\bf R}_2$ actually determines the electronic bandstructure of the tube. In particular, if $n-m=3k$ ($k$ being an integer) the tube is metallic while it is semiconducting or insulating otherwise\cite{Mintmire}.  A lot of discussions address the intrinsic state of metallic SWCNT where the Tomonaga Luttinger liquid state (TLL) may presumably occur\cite{Kane-PRL,TLL,Bockrath,Ishii,LL-recent}. In contrast to semiconducting nanotubes, where a general consensus is achieved\cite{Dressel}, unconventional features
of the metallic nanotubes are not well understood yet. A lot of discussions\cite{Kane-PRL,TLL,Bockrath,Ishii,LL-recent} address the linear dispersion law $\varepsilon _{k}=\pm v_{\mathrm{F}}\left\vert \mathbf{k}\right\vert $ [where "$\pm $" corresponds to electrons (holes), and $v_{\mathrm{F}}$ is the Fermi velocity], strong correlation effects, and a one dimensional transport of the electric charge carriers. Along with the TLL state in metallic SWCNT\cite{Kane-PRL,TLL,Bockrath,LL-recent,Ishii} under current attention are models of non-interacting electrons and of interaction with external environment\cite{Nazarov}. Despite indications of the TLL state in the shot noise\cite{LL-recent} and in angle integrated photoemission measurements\cite{Ishii}, present experimental evidences are still indirect\cite{Sonin,Nazarov}. This requires clearer identification of the intrinsic state in the one-dimensional quantum wells formed of metallic carbon nanotubes. 
Typical quantum well setup is sketched in Fig.~\ref{fig:Setup_b} where the 1D section is denoted as $%
\mathcal{T}$. The bias voltage $V_{\mathrm{ec}}$ drops between the emitter ($%
\mathcal{E}$) and collector ($\mathcal{C}$) electrodes, while the gate
voltage $V_{\mathrm{G}}$ is applied to the $n$%
++Si substrate as shown in Fig.~\ref{fig:Setup_b}. The electrochemical potentials in $%
\mathcal{E}$, $\mathcal{T}$ and $\mathcal{C}$ are denoted as $\mu _{e,T,c}$.
The $\mathcal{E}$ and $\mathcal{C}$ electrodes are separated from the
metallic nanotube section $\mathcal{T}$ by the interface barriers I shown in
black in Fig.~\ref{fig:Setup_b}(a,b). The potential barriers emerge from differences between
the Fermi velocities in the adjacent electrodes.

In this paper we suggest a method of identifying the quantized levels of charge and spin excitations in the Tomonaga Luttinger liquid state inside the 1D quantum well  shown in Fig.~\ref{fig:Setup_b}(a-c). The proposed here method exploits the fact that the electromagnetic field (EF) interacts with the charge excitations only, while the neutral particles remain unaffected. When tunneling electron with an energy $\varepsilon$ absorbs $n$ photons of EF, the multiphoton process probability depends on the intrinsic structure of the TLL state in $\mathcal{T}$. The PAT strongly influences the single electron tunneling (SET) and the level quantization taking place inside the quantum well which helps to elucidate its intrinsic state. In this paper we address the quantum well with a long and a short ${\cal T}$ section. When the ${\cal T}$ section is long, the quantization of electron motion inside ${\cal T}$ is neglected. Then the local single electron density of states ${\cal N}(\varepsilon )$ has a dip at zero energy $\varepsilon = 0$ inside ${\cal T}$. We will see that such a dip is well pronounced in the photon-assisted and in the single electron tunneling characteristics of the nanotube junction, which can be exploited for identification of the TLL state. Another type of QW corresponds to a short ${\cal T}$ section where the ballistic motion of the charged and neutral excitations inside the SWCNT section is quantized. During the tunneling, an electron splits into four $\rho _{\pm },\sigma _{\pm }$-bosons (two density and two spin). The bosons populate quantized levels with different energies $\varepsilon _{\rho +}\neq \varepsilon _{\rho -(\sigma \pm )}$. The charge boson energy levels are detected with the photon-assisted tunneling\cite{My-PRB} (PAT). Since the tunneling mechanism is sensitive to $V_{\mathrm{ec}} $ and $V_{\mathrm{G}}$, the TLL properties are therefore pronounced in the differential conductivity curves of the QW.   In the same setup, the spin boson levels are fingered from Zeeman splitting $\propto \mu_{\rm B} H$ when applying a finite d.c. magnetic field $H \neq 0$. The quantization of both the charge and spin excitations is pronounced in the differential conductivity curves of the quantum well. 

\section{Photon-assisted tunneling into the TLL state}
Here we address weakly coupled low-transparent double barrier SWCNT junctions, assuming that tunnelings across the $\mathcal{E/T}$ and $\mathcal{T/C}$ barriers are not phase-correlated. When the external electromagnetic field is polarized with the electric field vector directed along the nanotube axis, it induces an a.c. bias voltage $V^{(1)}$ across the whole double-barrier junction. The a.c. bias voltage effectively drops on the interface barriers I, which partial resistance is assumed to be much higher than the resistance of ${\cal T}$. Since a typical length of the carbon nanotube junction\cite{Paola-A,Paola-B} is $L \sim 200$~nm - $0.5$ $\mu$m, the a.c. field wavelength of interest is $1~{\rm mm} \geq \lambda_{\rm EF} \geq 0.5~\mu$m. This corresponds to the THz domain diapason. We describe tunneling between the TLL state in $\mathcal{T}$ and the free-electron states in $\mathcal{E}$ ($\mathcal{C}$)-electrode using the microscopic methods\cite{Emery,Keldysh,Datta}. For the sake of simplicity we do not consider here the ratchet effect\cite{Feldman,Trushin}, which comes either from an asymmetric scattering potential\cite{Feldman}, or from a nonlinearity of the electronic dispersion\cite{Trushin}. Using methods of Refs. \cite{Keldysh,Datta,Emery} one finds (see derivation details in Appendix) the time-averaged electric current through the quantum dot as%
\begin{eqnarray}
I &=&\Gamma _{\mathrm{n}} \frac{2e}{h}\int
d\varepsilon \sum_{nm}\zeta _{n}\mathrm{Im}{\cal K} \left(
\varepsilon \right) \nonumber  \\
&&\lbrack J_{m}^{2}\left( \alpha _{e}\right) \left[ f_{+}^{e}+{\cal G}\left( f_{-}^{e}-f_{+}^{e}\right) \right]  \nonumber \\
&&-J_{m}^{2}\left( \alpha _{c}\right) \left[ f_{+}^{c}+\left(
f_{-}^{c}-f_{+}^{c}\right) {\cal G}\right] ]
\label{I_PAT_SET}
\end{eqnarray}%
where $\mathcal{K}$ is the  full electron correlator, $f^{\mathrm{e(c)}}_{\pm }$ are the electron distribution functions in the $\mathcal{E}$ ($\mathcal{C}$)-electrodes for which we used a short notation $f\left(\varepsilon _{e\left( c\right)}^{m\pm }\right) \rightarrow f_{\pm }^{e\left( c\right) }$
\begin{eqnarray}
\varepsilon _{e\left( c\right) }^{m+} &=&\varepsilon +E+\hbar m\Omega
+\Delta U_{n}^{e,c} \nonumber \\
\varepsilon _{e\left( c\right) }^{m-} &=&\varepsilon +E+\hbar m\Omega
+\Delta U_{n-1}^{e,c},
\label{energ}
\end{eqnarray}
where  $E$ is the energy of the occupied level in the well relative to the conductance band edge in the emitter at the zeroth d.c. bias voltage $V_{\rm ec}=0$. In Eq.~(\ref{I_PAT_SET}) $\Gamma ^{c(e)}=4\pi e^{2}\nu _{e\left( c\right) }R_{e\left( c\right) }$, $\nu _{e\left( c\right) }$ is the electron density of states inside the $\mathcal{E}$($\mathcal{C}$), $R_{e\left( c\right) }$ is the tunnel resistance between the $\mathcal{E}$ ($\mathcal{C}$) electrodes and the $\mathcal{T}$ section, $\Gamma _{n}=\Gamma ^{e}\Gamma ^{c}/(\Gamma ^{e}+\Gamma ^{c})$, $\zeta _{n}$ is the probability to find $n$ electrons inside the well determined by a master equation (see Appendix), and the integration is performed over the electron energy $\varepsilon $ in the well. The Bessel function $J_{m}\left( \alpha _{e,c}\right) $ of order $m$ [$m$ is the number of emitted (absorbed) photons] in Eq.~(\ref{I_PAT_SET}) depends on the a.c. bias parameter $\alpha_{e,c}=eV_{e,c}^{(1)}/\hbar \Omega $ where $eV_{\mathrm{ec}}^{(1)}$ is the a.c. bias amplitude on the $\mathcal{E(C)}$
barriers, $\varepsilon$ is the electron energy in ${\cal T}$. The corresponding changes of emitter and collector electrostatic energy $\Delta U_{n}^{e,c}$ depend on the number $n$ of electrons in the well as%
\begin{eqnarray}
\Delta U_{n}^{e} &=&\delta \left( n+\frac{1}{2}\right) -\eta eV_{%
\mathrm{ec}} \nonumber \\
\Delta U_{n}^{c} &=&\delta \left( n+\frac{1}{2}\right) +\left(
1-\eta \right) eV_{\mathrm{ec}}
\label{U_en}
\end{eqnarray}%
where $\delta =e^2/C$, $C=C_{e}+C_{c}$ is the net capacitance, $C_{e\left( c\right) }$ is the
emitter (collector) capacitance, $\eta $ is the fraction of the net d.c. bias voltage $V_{ec}$, so that $\eta V_{ec}$ drops between the emitter and CNT. The electron distribution function ${\cal G}(\varepsilon )$ inside the tube entering Eq.~(\ref{I_PAT_SET}) must be, generally speaking, obtained from a corresponding quantum kinetic equation\cite{Keldysh}. However, to simplify our description of SET we will follow to Ref. \cite{Averin}. Namely we approximate ${\cal G}(\varepsilon )$ by a Fermi-Dirac distribution, but with a finite chemical potential $\mu_{\cal T} \neq 0$ in the form
\begin{equation}
{\cal G}_0(\varepsilon)=\frac{1}{\exp \left[ \left( \varepsilon -\mu_{\cal T} \right)
/T\right]+1}
\label{G_0}
\end{equation}
A similar quasiequilibrium approximation had formerly been used for describing of non-equilibrium superconductors\cite{Elesin} and of SET effects in semiconducting quantum wells. In Eq.~(\ref{G_0}) the chemical potential $\mu_{\cal T} $ of electrons in ${\cal T}$ is defined by the expression for the mean number of electrons
\begin{eqnarray}
\left\langle n\right\rangle  &=&\frac{2}{h}\frac{1}{\Gamma
^{e} +\Gamma ^{c}} \int d\varepsilon
\sum_{nm}\zeta _{n}\mathrm{Im}{\cal K}\left( \varepsilon
\right)  \nonumber \\ 
&&[\Gamma ^{e} J_{m}^{2}\left( \alpha
_{e}\right) \left[ 1-2f_{-}^{e}+2\left( f_{-}^{e}-f_{+}^{e}\right) {\cal G}\left( \varepsilon \right) \right]   \nonumber \\
&&+\Gamma ^{c}J_{m}^{2}\left( \alpha _{c}\right) \left[ 1-2f_{-}^{c}+2\left(
f_{-}^{c}-f_{+}^{c}\right) {\cal G}\left( \varepsilon \right) \right] ] \nonumber \\
&&+ \left\langle n_{\rm G}\right\rangle 
\label{n_ave}
\end{eqnarray}%
where $\left\langle n_{\rm G}\right\rangle$ is the number of extra electrons induced by the gate voltage $V_{\rm G} \neq 0$ applied as shown in Fig.~\ref{fig:Setup_b}. When the a.c. bias is off, $\alpha _{\mathrm{e,c}%
}\rightarrow 0$, Eq.~(\ref{I_PAT_SET}) yields well-known formula for the
electric conductivity of a double-barrier low transparent tunneling junction%
\cite{Averin}. In equilibrium and in absence of SET the distribution function ${\cal G}(\varepsilon )$ vanishes in Eq. (\ref{I_PAT_SET}) while $f^{e(c)}_{\pm }(\varepsilon )=f_{0}(\varepsilon )=1/(\exp{[\varepsilon/T]}+1)$, $T$ is the temperature. Then one simply gets
\begin{equation}
\sigma =(2e^{2}/h) \Gamma _{n} \overline{N},
\label{sigma_0}
\end{equation}
where $\overline{N}=\int_{-\infty}^{\infty } d\varepsilon \mathcal{N}\left(\varepsilon \right)\left(-  \partial f_{0}(\varepsilon )/(\partial \varepsilon)\right)$ is the number of conducting channels and $\mathcal{N}\left(\varepsilon \right)$ is the single electron density of states. The comb-shaped free electron density of states in the well is 
\begin{equation}
\mathcal{N}\left( \overline \epsilon \right) =\frac{1}{\pi}\mathrm{Im}\sum_{m}\frac{1}{\overline  \epsilon -m-i\zeta }
\end{equation} 
where $\zeta \rightarrow +0$ and we introduced the dimensionless energy $\overline \epsilon =\varepsilon /\Delta$, $\Delta =hv_{\rm F}/L$ is the level spacing, $h$ is the Plank constant. 
\begin{figure}[tbp]
\includegraphics{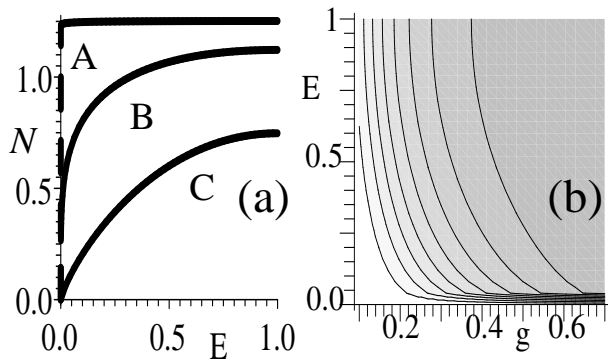}
\caption{(a)~The single electron density of states inside the $\mathcal{T}$
section of an QW for the Luttinger parameter $g=1$ (curve~A), $0.4$ (curve~B), and $g=0.2$. (b)~Contour $E-g$-plot of the single electron DOS inside
the $\mathcal{T}$~section.}
\label{fig:TLL_DOS}
\end{figure}

\section{The TLL tunneling density of states of a long QW}
If the metallic nanotube section\ $\mathcal{T}$\ is long, $L > v_{\mathrm{F}%
}\tau _{\mathrm{T}}$ (where $\tau _{\mathrm{T}}=\hbar /\Gamma _{n}$ is the
net tunneling time), the quantization inside $\mathcal{T}$ is negligible.
 The level separation for typical carbon nanotube junctions\cite{Paola-A,Paola-B} becomes indistinguishable when $L \geq 3$ $\mu $m. Strong electron correlations drive the electron system into the \textit{Tomonaga} \textit{Luttinger liquid} (TLL) state\cite{Kane-PRL, Bockrath,LL-recent}. According to Eq.~(\ref{I_PAT_SET}), the electric current is expressed via the single electron correlator $\cal K$, which is related to the spectral density $\mathcal{A}_{1}\left(\varepsilon\right)$ of the right-moving ($\kappa =1$) fermions as 
\begin{equation}
\mathcal{A}_{1}\left(\varepsilon\right) =\mathrm{Im}%
G_{1}^{R}\left( q=0,\varepsilon+i\delta \right) =-\mathrm{Im}\mathcal{K}\left(\varepsilon + i\delta \right) /\pi 
\label{A_QW}
\end{equation}
Following to Ref. \cite{Emery} at zero temperature $T=0$ one finds
\begin{eqnarray}
\mathcal{K}\left(\varepsilon \right) =-( 2/\sqrt{%
2\pi }) i|\varepsilon |^{2\gamma }\sin (\pi \gamma ) \nonumber \\ \cdot [2(-1)^{\gamma }\gamma \Gamma (-2\gamma -1)|\varepsilon | +e^{i\pi \gamma }{\rm sign}%
(\varepsilon ) \nonumber \\ 
( \Gamma (-2\gamma )+g^{2}( r^{2}/v_{F}^{2}) 
\cdot \gamma (2\gamma +1)\Gamma (-2(\gamma +1))\varepsilon^{2}) ],
\label{K_QW}
\end{eqnarray}
where $\Gamma (x)$ is the gamma function of $x$, $r$ is
the cutoff parameter, $\gamma =\left( g^{-1}+g-2\right) /8$, and $g$ is the
Luttinger liquid parameter. In the limit $g\rightarrow 1$, the expression
for $\mathcal{A}_{\kappa}\left(\varepsilon \right) $ transforms to the free
electron spectral density $\mathcal{A}_{\kappa}^{\left(0\right) }\left(\varepsilon \right) 
=\mathrm{sign}\varepsilon /\left( \sqrt{2\pi }v_{F}\right) $. Properties of
the TLL state are sensitive to the Luttinger parameter $g$. The single electron density of states ${\cal N} \left( E\right) $ of an QW with a long $\mathcal{T}$-section is shown for different values of $g$ in Figs.~\ref{fig:TLL_DOS}(a,b). The Luttinger parameter $g$ can be controlled either with the gate voltage $V_{\rm G}$ or with the split gate voltage $V_{\rm SG}$ as shown in Fig.~\ref{fig:Setup_b}(c) [see also the inset there].
\begin{figure}
 \includegraphics{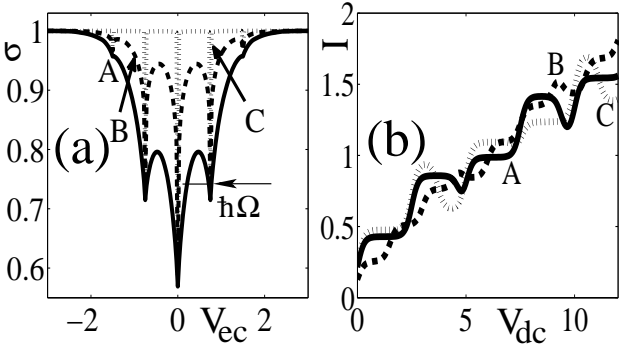}
 \caption{\label{fig:V_shape_TLL}
(a) Splitting of zero-bias TLL dip in the differential tunneling conductivity $\sigma (V_{\rm ec})$ [in units of $e^2 v_F N(0)$] in a long CNT junction due to the photon-assisted tunneling. Spacing between the zero dip and adjacent satellite dips is $\hbar \Omega$. (b)~The Coulomb staircase in the PASET current-voltage characteristics $I(V_{\rm dc})$ versus reduced voltage $V_{\rm dc}$ (see text) of the CNT junction in the TLL state with $g=0.4$ under influence of the a.c. bias field with amplitude $eV^{(1)}=3.4$ (in units of $\delta = e^2/C$) and for a symmetric junction ($\eta = 0.5$). Curve A corresponds to $\delta=5$ and $\Omega = 4.7$, curve B to $\delta=3$ and $\Omega = 4.7$ and curve C to $\delta=5$ and $\Omega = 3.7$.}
\end{figure}
Since the gate voltage $V_{\mathrm{G}}$ affects the charge density $q$ on $\mathcal{T}$, $q=CV_{\mathrm{G}}$ (where the capacitance $C=2\pi \epsilon _{0} L / \cosh^{-1}(2h_T/d)$, $\epsilon _{0}$ is the vacuum permittivity, $d$ is the nanotube diameter, and $h_T$ is the distance from the nanotube to substrate) it allows changing of $g$. An altering of $V_{\mathrm{G}}$ renormalizes $g\rightarrow g+\beta _{\mathrm{G}}V_{\mathrm{G}}$ due to changes in
the dielectric function $\epsilon (k,\varepsilon)$ and in the Coulomb screening. According to Refs. \cite{Kane-PRL,Egger}, the Luttinger parameter $g$ for a carbon nanotube depends on the electrostatic energy $U_n$ as
\begin{equation}
g \simeq \frac{1}{\sqrt{1+2U_n/\Delta }},
\label{g_Lut}
\end{equation}
where $\Delta $ is the energy level spacing while the change of $U_n$ is determined by Eq.~(\ref{U_en}).
A simple evaluation from the bandstructure calculations\cite{Lambert} gives $%
\beta _{\mathrm{G}}=0.005-0.03$ depending on directions of the rollup vector ${\bf R}$. The $\mathcal{N} \left( E\right) $ shape is also controlled with $V_{%
\mathrm{SG}}$ utilizing the split gate configuration\cite{Lambert} as shown
in Fig.~\ref{fig:Setup_b}(c). The electric field in that setup is perpendicular to the
nanotube axis as shown in the right inset to Fig.~\ref{fig:Setup_b}(c).  The split-gate setup allows to drive the tube electron state from the semiconducting to the metallic one.  The transversal electric field induces a finite
dipole momentum directed in perpendicular to the nanotube, which again renormalizes $g$. The corresponding alteration of the Luttinger parameter $g$ is evaluated using, e.g., results of Refs. \cite{Keeffe,Chen,Chiu}. This gives $g \rightarrow g\cdot J_{0}^{-1}\left( V_{\mathrm{SG}}d/\hbar v_{\mathrm{F}}\right) $. For a narrow-gap $\mathcal{T}$ (see Ref. \onlinecite{Lambert}) and for typical parameters of the nanotube quantum well, the split gate induced change is $g \rightarrow g+\beta_{\mathrm{SG}}V_{\mathrm{SG}}$ where $\beta _{\mathrm{SG}}=0.01-0.05$ for different rollup vectors. If the electric field $V_{\mathrm{SG}}/d$ is sufficiently strong, one induces a semiconducting-metal transition\cite{Keeffe,Chen,Chiu}. The electronic properties of the nanotube then switch from a one-dimensional semiconductor to the TLL.

The time-averaged conductance $\sigma (V_{\rm ec})$  of the long CNT junction exposed to an external electromagnetic field is computed using Eqs. (\ref{I_PAT_SET})-(\ref{U_en}), (\ref{n_ave}), (\ref{A_QW}), and (\ref{K_QW}). We calculated $\sigma (V_{\rm ec})$ and the electric current $I(V_{\rm dc})$ ($V_{\rm dc}=(V_{\rm ec}-V_t)\eta$ is the reduced voltage) for two cases of interest, $V_t=(E-\epsilon_{\rm F})/e\eta$ is the SET threshold voltage. One limit corresponds to  $\delta << T$ when the single-electron tunneling is not essential [$\delta =e^2/C$, $C=C_{e}+C_{c}$ is the net capacitance of the double-barrier junction, $C_{e\left( c\right) }$ is the emitter (collector) capacitance]. Then Eq.~ (\ref{I_PAT_SET}) for the tunneling current through the quantum well is simplified as
\begin{eqnarray}
I\left( V_{\mathrm{ec}}\right) &=&\frac{2e}{h}\Gamma
_{n}\sum_{km}\int d\varepsilon \mathrm{Im}{\cal K}\left(\varepsilon,\Omega \right) \nonumber \\
&& \cdot  [J_{m}^{2}\left( \alpha _{e}\right) f\left( \varepsilon
_{m}^{e}\right) -J_{m}^{2}\left( \alpha _{c}\right) f\left( \varepsilon _{m}^{c}\right) ]%
\label{current}
\end{eqnarray}%
where now
\begin{eqnarray}
\varepsilon _{m}^{e}=\varepsilon_{\rm ec} -\eta eV_{\rm ec}+\hbar m\Omega \nonumber \\
\varepsilon _{m}^{c}=\varepsilon_{\rm ec} +\left( 1-\eta \right) eV_{\rm ec}+\hbar m\Omega ,
\label{energ0}
\end{eqnarray}
The above Eqs.~(\ref{current}), (\ref{energ0}) are completed by Eq.~(\ref{A_QW}) to compute the d.c. differential tunneling conductance $\sigma(V_{ec})$. The results are shown in Fig.~\ref{fig:V_shape_TLL}(a) where we plot $\sigma (V_{\rm ec})$ of a long CNT junction in conditions of the photon-assisted tunneling  for $\Omega = 0.75$, $eV^{(1)}=0.65$ and for three different values of the Luttinger liquid parameters $g=0.2$ (curve A), $g=0.4$ (curve B), and $g=0.93$  (curve C). One may notice that the zero-bias dip in $\sigma (V_{\rm ec})$, which was positioned at $V_{\rm ec}=0$ when the a.c. field was off ($eV^{(1)}=0$) splits into additional satellite peaks spaced by $\hbar \Omega$. 

Another limit corresponds to the single electron tunneling which occurs if the condition 
\begin{eqnarray}
eV_{\mathrm{ec}}/2-\delta /2-E-\hbar m\Omega -\delta \cdot n\leq \mu_{\cal T}
\left( n\right) \leq \nonumber \\
eV_{\mathrm{ec}}/2+\delta /2-E-\hbar m\Omega -\delta
\cdot n  \label{SET_cond}
\end{eqnarray}%
is fulfilled. The condition (\ref{SET_cond}) can be independently accomplished by adjusting $%
\Omega $, $V_{\rm G}$, and $V_{\rm ec}$. The zero-temperature conductivity for a symmetric junction then takes the form%
\begin{eqnarray}
\sigma \left( V_{\mathrm{ec}}\right)  = \frac{\partial I\left( V_{\mathrm{%
ec}}\right) }{\partial V_{\mathrm{ec}}}=\frac{2e}{h}\Gamma _{ec} \sum_{m}J_{m}^{2}\left( \alpha \right) \zeta
_{n} \nonumber \\
\cdot \lbrack \mathrm{Im}{\cal K}\left( \mathcal{E}%
_{+}^{m}\left( n\right) \right)   
+\mathrm{Im}{\cal K}\left( \mathcal{E}_{-}^{m}\left( n\right) \right)
\cdot \theta \left( \mu_{\cal T} \left( n\right) -\mathcal{E}_{-}^{m}\left( n\right)
\right)   \nonumber \\
-\mathrm{Im}{\cal K}\left( \mathcal{E}_{+}^{m}\left( n\right) \right)
\cdot \theta \left( \mu_{\cal T} \left( n\right) -\mathcal{E}_{+}^{m}\left( n\right)
\right) ]
\label{sig_A}
\end{eqnarray}%
where 
\[
\mathcal{E}_{\pm }^{m}\left( n\right) =eV_{\mathrm{ec}}-E-\hbar m\Omega
-\delta \left( n\pm \frac{1}{2}\right) 
\]%
and $E$ is the occupied level energy in the quantum well relative to the
conductance band edge in the emitter in absence of the bias voltage. Eq.~(\ref{sig_A}) can be overwritten in the shorter form%
\begin{equation}
\sigma \left( V_{\mathrm{ec}}\right) =\frac{2e}{h}\Gamma _{ec} \sum_{m}J_{m}^{2}\left( \alpha \right) \zeta
_{n}\cdot \mathcal{A}\left( n,m\right)   \label{sig_B}
\end{equation}%
where 
\[
\mathcal{A}\left( n,m\right) =\left \{
\begin{array}{c}
\mathrm{Im}{\cal K}\left( \mathcal{E}_{+}^{m}\left( n\right) \right) 
\text{ if }\mu_{\cal T} \left( n\right) <\mathcal{E}_{+}^{m}\left( n\right)  \\ 
 0\text{ \ \ \ \ if }\mathcal{E}_{+}^{m}\left( n\right) <\mu_{\cal T} \left(
n\right) <\mathcal{E}_{+}^{m}\left( n\right)  \\ 
\mathrm{Im}{\cal K}\left( \mathcal{E}_{-}^{m}\left( n\right) \right) 
\text{ if }\mu_{\cal T} \left( n\right) >\mathcal{E}_{-}^{m}\left( n\right) 
\end{array}\right.
\]%
since one always gets $\mathcal{E}_{+}^{m}\left( n\right) <\mathcal{E}_{-}^{m}\left(
n\right) $. To consistently describe the single electron tunneling Eqs.~(\ref{I_PAT_SET})-(\ref{U_en}) must be completed by an equation for $%
\mu_{\cal T} \left( n\right) $. For a symmetric junction ($\alpha_e=\alpha_c$) one gets 
\begin{eqnarray}
\left\langle n_{\mathrm{G}}\right\rangle +\left\langle n\right\rangle  &=&\frac{4}{h}\cdot \sum_{n,m}J_{m}^{2}\left( \alpha \right)
\zeta _{n}\int d\varepsilon   \nonumber \\
&&\mathrm{Im}{\cal K}\left( \varepsilon \right) [1-f_{-}^{ec}+\left(
f_{-}^{ec}-f_{+}^{ec}\right) {\cal G}\left( \varepsilon \right) ]
\nonumber
\end{eqnarray}%
where
\begin{eqnarray}
f_{\pm }^{ec} &=&f\left( \varepsilon +E+\hbar m\Omega +\delta \left( n\pm 
\frac{1}{2}\right) -eV_{\mathrm{ec}}/2\right)   \nonumber \\
&&+f\left( \varepsilon +E+\hbar m\Omega +\delta \left( n\pm \frac{1}{2}%
\right) +eV_{\mathrm{ec}}/2\right) 
\nonumber
\end{eqnarray}%
and ${\cal G}\left( \varepsilon \right)$ is approximated by Eq. (\ref{G_0}), $n_{\mathrm{G}}$ is the additional electron density induced by a finite gate voltage $V_{\mathrm{G}}\neq 0$. The condition which determines a vertical step in the $I(V_{ec})$ [or a sharp peak in $\sigma (V_{ec})$] is 
\[
e\eta \left( V_{k-1,k}^{m}-V_{t}\right) =\delta \left( k-1\right)
+\varepsilon +m\hbar \Omega
\]%
where $k$ is integer. At $\alpha_{e,c}=0$ one gets
\[
V_{k-1,k}=V_{t}+\frac{\delta \left( k-1\right) +\varepsilon }{e\eta }
\]%
The spacing between two adjacent steps at $\alpha _{e,c}=0$ is%
\begin{eqnarray}
V_{k,k+1}-V_{k-1,k} =\left( V_{t}+\frac{\delta k+\varepsilon _{k+1}}{e\eta 
}\right) \nonumber  \\
-\left( V_{t}+\frac{\delta \left( k-1\right) +\varepsilon _{k}}{%
e\eta }\right)  
= \frac{\delta +\varepsilon _{k+1}-\varepsilon _{k}}{e\eta }
\end{eqnarray}%
When the external a.c. field is finite ($\alpha _{e,c}\neq 0$), the steps are
splet additionally by $\pm \hbar \Omega $.

The current-voltage characteristics $I(V_{\rm dc})$ in condition of the single-electron photon-assisted tunneling (PASET) across the quantum well in the Tomonaga Luttinger liquid state are shown in Fig.~\ref{fig:V_shape_TLL}(b). According to Ref. \onlinecite{Averin}, equilibrium shape of  the $I(V_{\rm dc})$ curves (quoted as Coulomb staircase) is extremely sensitive to the double-barrier junction's parameters such as barrier transparencies, capacitance, symmetry,  purity of the carbon nanotube section, and the energy level spacing. The photon-assisted tunneling induced by the external electromagnetic field introduces additional features in those curves. We compute PASET curves for a QW with a long ${\cal T}$--section where the single electron tunneling takes place. Remarkable elements of the $I(V_{\rm dc})$ curves A-C in Fig.~\ref{fig:V_shape_TLL}(b) are local dips which originate from an interference between the zero-energy TLL anomaly pronounced in equilibrium at $\varepsilon = 0$ [see Fig.~\ref{fig:TLL_DOS}(a)] and the photon-assisted single electron tunneling processes. The Coulomb staircase curve A in Fig.~\ref{fig:V_shape_TLL}(b) corresponds to $\delta=5$ and $\Omega = 4.7$, curve B to $\delta=3$ and $\Omega = 4.7$ and curve C to $\delta=5$ and $\Omega = 3.7$ computed for $g=0.4$. 

\section{Recognition of charge and spin boson energy levels}
\begin{figure}
 \includegraphics{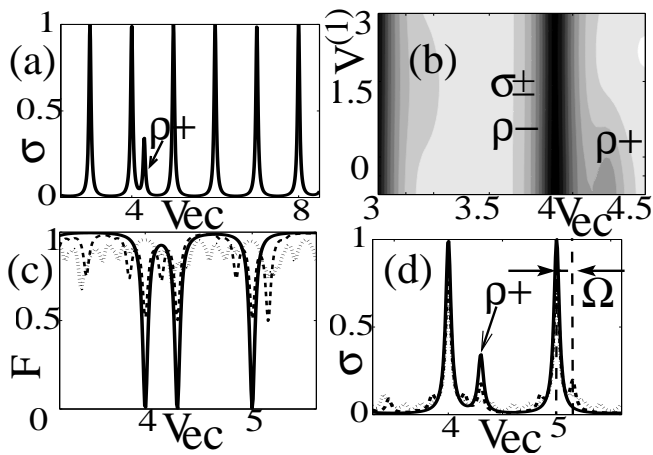}
 \caption{\label{fig:Combs_TLL}
(a) The steady state tunneling differential conductivity ${\sigma}(V_{\rm ec})$ [in units of $e^2 v_F N(0)$] in the TLL state. The peak at $V_{\rm ec}=4.3$ (in units of $\Delta/e$) corresponds to the $\rho _{+}$ -boson. (b) The contour plot $\sigma (V_{\rm ec},V^{(1)}_{e,c})$ [$V_{\rm ec}$ and the a.c. bias amplitude $V^{(1)}_{e,c}$ being in units of $\Delta/e$]. (c) The Fano factor $F(E)$ of the QW in conditions of PAT. (d) ${\sigma}(V_{\rm ec})$ for different $\Omega $.}
\end{figure}

\begin{figure}
 \includegraphics{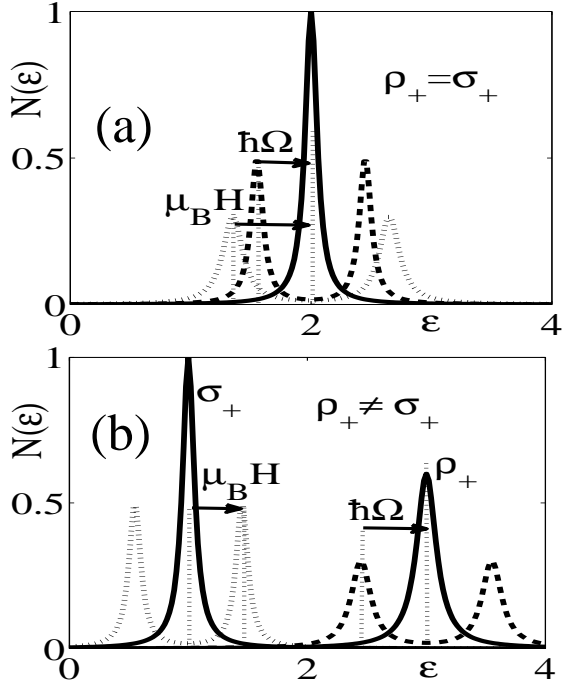}
 \caption{\label{fig:two_peak_TLL}
 Splitting of quantized levels due to the photon assisted tunneling and Zeeman effect as pronounced in the single electron density of states $N(\varepsilon)$ of a short CNT junction. (a)~A free electron quantized level (solid curve) for which charge $\rho_{+}$ and spin $\sigma_{+}$ bosons coincide splits by an a.c. electromagnetic field (with spacing $\propto \hbar \Omega$) and by a d.c. magnetic field (with spacing $\propto \mu_{\rm B} H$) simultaneously. (b)~Quantized levels of spin $\sigma_{+}$ and charge $\rho_{+}$ bosons have different energies $\varepsilon^{\rho +}_n \neq \varepsilon^{\sigma +}_n$ in the TLL state. The a.c. bias splits the charge boson levels (dashed curve on right) while the d.c. magnetic field splits the spin boson levels (dotted curve on left) only.
The charge boson localized energy level  $\varepsilon^{\rho+}_n$ splits in two satellite peaks with spacing $2\hbar \Omega$. Although the a.c. field has no influence to the neutral spin bosons, the spin level $\varepsilon^{\sigma+}_n$ splits\cite{Kane-PRL} in two sublevels spaced with $\Delta_{\rm Z} \propto \mu_{\rm B} H$  (both in units of $\Delta$) due to Zeeman effect when a d.c. magnetic field $H \neq 0$ is applied.}
\end{figure}

In a opposite limit when the ${\cal T}$--section is short, the quantized energy levels are well resolved, $\Gamma ^{e,c}<<\Delta $ ($\Gamma ^{e,c}$ are the $\mathcal{E}\Leftrightarrow 
\mathcal{T}$ and $\mathcal{T}\Leftrightarrow \mathcal{C}$ electron tunneling
rates, $\Delta = h v_{\rm F}/L$ is the interlevel spacing inside $\mathcal{T}$). In this Section we neglect the single-electron tunneling (SET) contribution\cite{Averin} (Coulomb blockade phenomena) which is justified when the temperature $T$ is not too low, $T>>\Gamma ^{e,c}$. In that limit we use Eq.~(\ref{current}) again but with a different ${\rm Im}{\cal K (\varepsilon)}$ which now has a comb-like shape. Due to the spin-charge separation in the Tomonaga Luttinger liquid (TLL) there are two sets of quantized energy levels in a low-transparent quantum well with a short ${\cal T}$--section. For the QW transparency $T=0.3$ (where $T=4\pi \Gamma_{n} L_{n}/(\hbar v_F)$, $%
L_{n}=L_{e}+L_{c}$, $L_{e}$ and $L_{c}$ are the $\mathcal{E}$ and $\mathcal{C%
}$ thicknesses respectively, $v_F=8.1 \cdot 10^5$ m/s) one gets $\Gamma
_{n}\approx 0.3$ meV. For the nanotube length $L = 3~\mu$m one obtains
spacing between the quantized levels as $\Delta = 1$ meV. The photon-assisted
processes cause an additional splitting $\sim $ 0.6 meV which corresponds to the a.c. bias frequency $\Omega \approx 1$ THz. Following to Refs. \cite{Kane-PRL,Datta}, the transmission coefficient is $T (E) = \vert i \hbar G^R \left( L,E\right)\vert^2$. We assume that coupling of the single wall nanotube segment ${\cal T}$ to the external ${\cal E}$ and ${\cal C}$ electrodes is weak. In this approximation we compute the local electron density of states ${\cal N}(\varepsilon )$ implementing boundary conditions\cite{Kane-PRL} for the electron wavefunction inside a short  carbon nanotube section ${\cal T}$. Then the quantized energy levels are well separated and resolved. The retarded single electron Green function is $G^R\left( L,t\right) =\Pi _{a}\mathcal{G}^R_{a}\left( L,t\right) $, which Fourier transform has a comb-like shape 
\begin{eqnarray}
G^R_n\left(L,\varepsilon \right) =i%
\sqrt{\frac{2}{\pi}}\sum_{a} \frac{4^{g_{a}^{-}} \sin ^{2g_{a}^{-}}\left( \pi \lambda /L\right)}{\varepsilon_{a}} 
\nonumber \\
\sum_{n}\Theta _{n}\left(-1\right) ^{-2g_{a}^{+}+n} 
\cdot \frac{\Gamma \left( 1-2g_{a}^{+}+n\right) \Gamma \left(\varepsilon /\varepsilon _{a}\right) }{
 \Gamma \left( 1-2g_{a}^{+}+n+\varepsilon /\varepsilon
_{a}\right) }
\label{g_R}
\end{eqnarray} 
where $a=\left( \rho_\pm ,\sigma_\pm \right) $ is
the TLL boson index, i.e., $\varepsilon^{\rho +}_n = \Delta/g$ while $%
\varepsilon^{ \rho - (\sigma \pm)}_n = \Delta$, $\Theta _{n}$ are the
coordinate-dependent coefficients inside the nanotube and the length parameter $\lambda <<L$
effectively incorporates influence of the interface barriers\cite{Kane-PRL}, $n$ is the
quantization index, and $g_{a}^{\pm}=(1/g_a \pm g_a)/16$. The parameters $\Theta _{n}$ and $\lambda$ are determined by the integer charge and by the sum of phase shifts at the interfaces.  The charge $\rho _{+} $ bosons populate the energy levels $\varepsilon_{n}^{\rho +}=hv_{F}n/Lg = n \Delta /g$ (where $n$ is
integer number), while three other neutral $\rho _{-}$ and $\sigma _{\pm
}$-boson energy levels have conventional values $\varepsilon_{n}^{\rho -(\sigma
\pm) }=hv_{F}n/L=n\cdot \Delta$. The tunneling differential
conductivity $\sigma (V_{\mathrm{ec}})$ of a "clean" (i.e., without impurities on $\mathcal T$) sample is a combination of two combs with different periods shown in Fig.~\ref{fig:Combs_TLL}(a) for $g=0.23$. One of the combs corresponds to the $\rho _{+}$-boson, while another comb is related to  three remaining neutral $\left( \rho _{-},\sigma _{+},\sigma _{-}\right) $-bosons. 
In the steady state when the EF is off (i.e., $\alpha _{\rm ec}\equiv 0$), during the tunneling say, from $\mathcal{E}$ to $\mathcal{T}$, an electron splits into four bosons as $e\rightarrow \rho _{+}+\rho _{-}+\sigma _{+}+\sigma _{-}$, which assumes the energy conservation as 
\begin{eqnarray}
E+\eta eV_{\rm ec}=\varepsilon_{n}^{\rho +}+\varepsilon _{n}^{\rho -}+\varepsilon _{n}^{\sigma +}+\varepsilon _{n}^{\sigma -} \nonumber \\
=n \left(3+1/g\right) \Delta
\end{eqnarray}
 ($n$ being the integer number). That corresponds to a resonance tunneling through the quantized TLL states tuned by $V_{\rm ec}$. However, if the electromagnetic field (EF) is on ($\alpha _{\rm e,c}\neq 0$), the resonance tunneling condition changes. That happens because the a.c. field acts on the charge $\rho _{+}$-bosons only, which absorb the EF photons during the photon-assisted tunneling processes. The photons do not excite the neutral $\rho _{-}$ and $\sigma _{\pm }$ bosons since they do not interact with the EF. The EF-modified resonance condition depends on both $V_{\rm ec}$ and $\Omega $ simultaneously%
\begin{eqnarray}
E+\eta eV_{\rm ec} &=&\left( \varepsilon _{n}^{\rho +}+m\hbar \Omega \right)
+ \varepsilon _{n}^{\rho -} +\varepsilon
_{n}^{\sigma +}+\varepsilon _{n}^{\sigma -}  \nonumber \\
&=& n \left( 3+1/g\right) \Delta+m\hbar \Omega 
\label{ac_reson}
\end{eqnarray}%
where $n$  and $m$  are integer numbers.
Since the electromagnetic field acts on the charge bosons only (not affecting the neutral bosons at all) it splits the corresponding peaks selectively.  Because $V_{\rm ec}$ and $\Omega $ are bound by the condition (\ref{ac_reson}), this imposes a constrain on the net photon-assisted tunneling (PAT) resonant current through the quantum well. Using Eq.~(\ref{ac_reson}) one immediately extracts $g$ and $\Delta$ from the PAT current-voltage characteristics. More specifically, the value of $g$ and $\Delta$ follow right from periods of the two steady state combs $\sigma (V_{ec})$ shown in Fig.~\ref{fig:Combs_TLL}(a). This is illustrated by the PAT differential tunneling conductivity $\sigma(V_{\rm ec})$ for $\alpha_{\rm e,c} \neq 0$ shown in Figs.~\ref{fig:Combs_TLL}(b,d). Fig.~\ref{fig:Combs_TLL}(b) is the contour plot  $\sigma (V_{\rm ec},V^{(1)}_{e,c})$ ($V_{\rm ec}$ and the a.c. bias amplitude $V^{(1)}_{e,c}$ being in units of $\Delta/e$). The same quantity $\sigma(V_{\rm ec})$ but for fixed $\Omega = 0.85$ and $eV^{(1)}_{\rm e,c}=0.01$ (solid curve), $eV^{(1)}_{\rm e,c}=1$ (dashed curve), and $eV^{(1)}_{\rm e,c}=5$ (dotted curve) is presented in Fig.~\ref{fig:Combs_TLL}(d). In an experiment one obtains series of peaks in the differential conductivity $\sigma(V_{\rm ec})=\partial I /\partial V_{\rm ec}$ curves for the steady state ($\alpha _{\rm e,c}\equiv 0$) and the PAT peaks when the EF is on ($\alpha _{\rm e,c}\neq 0$). Then one determines the ratio $r_1=A_1^{\rho +}/A_0^{\rho +}$ where $A_{0(1)}^{\rho +}$ is the $\rho_{+}$-boson peak height, which corresponds to the number of emitted (absorbed) photons $m=0,1$. The ratio $r_1$ allows extracting of the actual a.c. field amplitude $V^{(1)}$ acting on the junction. Since the a.c. field acts on the $\rho_{+}$ bosons only, the EF helps to identifying of the TLL state. The method is illustrated further in Fig.~\ref{fig:two_peak_TLL} where we show a single peak in ${\cal N}(\varepsilon )$ corresponding to a quantized free electron energy level [see Fig.~\ref{fig:two_peak_TLL}(a)]. For non-interacting electrons ($g=1$) the same single level splits either by an a.c. field due to the photon-assisted tunneling phenomena with spacing $\propto m\hbar \Omega$ ($m$ being integer) or by a d.c. magnetic field with the Zeeman spacing $\propto \mu_{\rm B} H$. The situation is remarkably different in the Luttinger liquid state when $g \neq 1$ and the charge $\rho_{+}$ and spin $\sigma_{+}$ levels have different energies $\varepsilon^{\rho+}_n \neq \varepsilon^{\sigma+}_n$. Then one easily identifies the charge and spin levels merely by applying the a.c. field and d.c. magnetic field to the same quantum well.  If a level splits with spacing $\propto m\hbar \Omega$ by the a.c. field only (showing no response to the d.c. field) then it certainly is a $\rho_{+}$ charge boson level ($g \neq 1$). If it splits by the d.c. magnetic field\cite{Kane-PRL} with the Zeeman spacing $\propto \mu_{\rm B} H$ showing no response to the a.c. field, then it must be associated with the spin bosons $\sigma_{+}$. However if the both a.c. and d.c. magnetic fields split the same level, then the level belongs to the non-interactive electrons ($g=1$) as had been noticed above. In this way one perceives the charge and spin bosons in experiments when applying a.c. electromagnetic field in combination with the d.c. magnetic field to a carbon nanotube junction. An important requirement to  the experimental metallic carbon nanotube quantum well samples is that they must be clean.  An electron-impurity scattering in real samples leads to a formation of additional pairs of combs with different periods. Then, an identification of the TLL  state  becomes possible with a mere generalization of the method described above. 

Ratio of the noise power to the mean current (Fano factor) is computed as $F=\sum_n T _n (1-T _n)/\sum_n T _n$. The result is shown in Fig.~\ref{fig:Combs_TLL}(c). One can see that at the energies of quantized levels the noise is much lower than the Poisson noise of a conventional tunnel  junction where $F=1$.  Remarkably, the multiphoton absorption  is pronounced in the noise spectra as well. In this way Fig.~\ref{fig:Combs_TLL}(c) suggest a method of the noise spectroscopy for studying of the photon-assisted tunneling into the TLL state.

\section{Conclusions}
Phenomena considered in this paper originate from a specific physics of the
charge and spin carriers, behaving like a blend of four non-interacting bosons. The
Tomonaga Luttinger liquid state occurs inside the one dimensional quantum well formed by a metallic single wall carbon nanotube. The TLL state is tested with applying of an external a.c. electromagnetic field and of a d.c. magnetic field simultaneously. The a.c. field splits the charge boson energy levels due to the photon-assisted tunneling while the d.c. magnetic field splits the spin boson levels due to the Zeeman effect. That allows a mere identification of the quantized energy levels associated with the charge and spin bosons forming the TLL state in relevant experiments. Besides one also finds the quantized level spacing $\Delta$ and the TLL parameter $g$. The unconventional electronic and photonic properties of the metallic carbon nanotube quantum well can be utilized in various nanodevice applications including THz field sensors and nanoemitters.

I wish to thank V. Chandrasekhar and P. Barbara for fruitful discussions.
This work had been supported by the AFOSR grant FA9550-06-1-0366.

\section{Appendix}
Here we derive analytical expressions for the time-averaged electric current
through the TLL quantum well in conditions of the  photon-assisted single
electron tunneling (PASET). The external electromagnetic field is applied
to the nanotube junction as shown in Fig.~\ref{fig:Setup_b}. Our model describes a low
transparency junction which average conductance is small, $G<<R_{Q}^{-1}$
where $R_{Q}=\pi \hbar /2e^{2}\simeq 6.5$~k$\Omega $. Since coupling to the
electrodes is weak, the emitter-nanotube ($\mathcal{E}\Leftrightarrow 
\mathcal{T}$) and the nanotube-collector ($\mathcal{T}\Leftrightarrow 
\mathcal{C}$) tunneling processes are assumed as being not phase-correlated.
The SET dynamics in that approximation is well described by a simple master
equation\cite{Averin}. The external electromagnetic field with frequency $%
\Omega $ induces an a.c. bias voltage with amplitude $V^{(1)}\cos \Omega t$ across the
junction. The a.c. voltage modulates phases of the tunneling electrons as $\phi
^{e,c} \left( t\right) =\left( eV^{(1)}_{e,c}/\hbar \right) \int^{t}\cos \Omega
t^{\prime }dt^{\prime }$ where $V^{(1)}_{e}=\eta V^{(1)}$, $V^{(1)}_{c}=\left(
1-\eta \right) V^{(1)}$ are corresponding fractions of the a.c. voltage drop
on the emitter and collector, $\eta $ is the fraction of the net a.c. bias
voltage $V^{(1)}$, so $\eta V^{(1)}$ drops between the emitter and CNT. 
The time-averaged single electron tunneling electric current through the double
barrier junction is expressed via partial tunneling rates $w_{+}^{e,c}$ and $w_{-}^{e,c}$ from emitter (collector) to the
energy levels inside the well\cite{Averin} 
\begin{eqnarray}
w_{+}^{e,c}\left( n,m\right)  = J_{m}^{2}\left( \alpha
_{e,c}\right) \int d\varepsilon \Gamma _{\varepsilon }^{e,c} \left(
\varepsilon _{e\left( c\right) }^{m+}\right)  \nonumber \\ 
\cdot  f\left( \varepsilon _{e\left(
c\right) }^{m+}\right) G_{m}^{-}\left( \varepsilon ,\Omega \right)  
\nonumber \\
w_{-}^{e,c}\left( n,m\right)  = -J_{m}^{2}\left( \alpha
_{e,c}\right) \int d\varepsilon \Gamma _{\varepsilon }^{e,c}\left(
\varepsilon _{e\left( c\right) }^{m-}\right) \nonumber \\ 
\cdot \left[ 1-f\left( \varepsilon
_{e\left( c\right) }^{m-}\right) \right] G_{m}^{+}\left( \varepsilon
_{k},\Omega \right)   \label{tunn_rates}
\end{eqnarray}%
\[
w_{nm}^{\pm }=w_{\pm }^{e}\left( n,m\right) +w_{\pm }^{c}\left( n,m\right) 
\]%
where the integration is performed over the electron energy $\varepsilon $
in the well, $E$ is the energy of the occupied level in the well relative to
the conductance band edge in the emitter at the zeroth d.c. bias voltage $V_{%
\mathrm{e}\left( \mathrm{c}\right) }=0$, $m$ is the number of emitted
(absorbed) photons, and the electron energy arguments are
\begin{eqnarray*}
\varepsilon _{e\left( c\right) }^{m+} &=&\varepsilon +E+\hbar m\Omega
+\Delta U_{n}^{e,c} \\
\varepsilon _{e\left( c\right) }^{m-} &=&\varepsilon +E+\hbar m\Omega
+\Delta U_{n-1}^{e,c}
\end{eqnarray*}%
The Bessel function $J_{m}\left( \alpha _{e,c}\right) $ of order $m$ in Eq.~(%
\ref{tunn_rates}) depends on the a.c. bias parameter $\alpha
_{e,c}=eV_{e,c}^{(1)}/\hbar \Omega $. Corresponding changes $\Delta U_{n}^{e,c}$ of the emitter and collector electrostatic energy depend on the number $n$ of electrons in the well as%
\begin{eqnarray*}
\Delta U_{n}^{e} &=&\delta \left( n+\frac{1}{2}\right) -\eta eV_{%
\mathrm{ec}} \\
\Delta U_{n}^{c} &=&\delta \left( n+\frac{1}{2}\right) +\left(
1-\eta \right) eV_{\mathrm{ec}}
\end{eqnarray*}%
where $\delta = e^{2}/C$, $C=C_{e}+C_{c}$ is the net capacitance, $C_{e\left( c\right) }$ is the
emitter (collector) capacitance. In Eqs.~(\ref{tunn_rates}), the electron Keldysh Green function\cite{Keldysh} $%
G^{\pm }$ in the well is defined in the $\left\{\mathbf{r},t\right\} $-presentation as%
\begin{equation}
G^{\pm }\left( \mathbf{r},\mathbf{r}^{\prime },t,t^{\prime
}\right) =\pm i\,\left\langle \psi \left( \mathbf{r},t_{\pm
}\right) \psi ^{\dagger }\left( \mathbf{r}^{\prime },t_{\mp
}^{\prime }\right) \right\rangle 
\end{equation}%
where $\mathbf{r}$, $\mathbf{r}^{\prime }$ are electron coordinates and $%
t_{\pm }$, $t_{\mp }^{\prime }$ are the time moments assigned to points
lying on the positive (+) or negative (-) branch of the contour $c$ circled
around the time axis $-\infty <t<\infty $, $\left\langle \ldots
\right\rangle $ means averaging\cite{Keldysh}  with full Hamiltonian ${\hat {\cal H}}$ which includes all the interactions in the system. Following to Ref. \onlinecite{Emery} one may introduce the auxiliary right-moving ($\kappa =1$) free fermion Green Keldysh function as 
\begin{equation}
G_{10}^{\pm }=\frac{i\pi T/v_{F}}{\sinh \left(
\pi T\left( x-v_{F}t\right) /v_{F}\right) }\pm i\pi \delta \left(
x-v_{F}t\right),
\label{G_10}
\end{equation}
 and for the left-moving ($\kappa =2$) free fermion $G_{20}^{\pm }=[G_{10}^{\pm }\{v_{\rm F} \rightarrow -v_{\rm F}\}]^{*}$. The Fourier transforms  of (\ref{G_10}) at $x=0$ are
\begin{eqnarray}
G_{10}^{+}\left(\varepsilon \right) =i\mathcal{N}\left(\varepsilon \right) \left(1-{\cal G}_0\left(\varepsilon \right) \right) \nonumber \\
G_{10}^{-}\left(\varepsilon \right)=-i\mathcal{N}\left(\varepsilon \right) {\cal G}_0\left(\varepsilon \right) 
\label{G_pm}
\end{eqnarray}
where $\mathcal{N}\left(\varepsilon \right)$ is the single electron density of states inside the well, ${\cal G}_0\left(\varepsilon \right)=(1-\tanh{\left(\varepsilon /2T\right)})/2$ is the equilibrium electron distribution function in the well.  The electron tunneling rate $\Gamma _{k}^{e,c}$ is modified by the
external a.c. bias as%
\begin{eqnarray*}
\Gamma ^{e,c}\left(t\right) \rightarrow \Gamma ^{e,c}\left(t\right) e^{i\alpha _{e,c} \cos
\Omega t} \\
=\int d\varepsilon ^{\prime }\Gamma ^{e,c}\left( \varepsilon
^{\prime }\right) e^{i\varepsilon ^{\prime }t/\hbar }e^{i\alpha _{e,c}\cos \Omega t}
\end{eqnarray*}
The backward Fourier transform gives%
\begin{eqnarray}
\Gamma ^{e,c}\left( \varepsilon
\right) &=&\int dt\int d\varepsilon ^{\prime }\Gamma ^{e,c}\left( \varepsilon
^{\prime }\right) e^{i\alpha \cos \Omega t}e^{-i\varepsilon t/\hbar+i\varepsilon
^{\prime }t/\hbar} \nonumber  \\
&=&\sum_{m}J_{m}^{2}\left( \alpha _{e,c}\right) \int d\varepsilon ^{\prime
}\int dt\Gamma ^{e,c}\left( \varepsilon ^{\prime }\right) e^{im\Omega
t-i\varepsilon t/\hbar+i\varepsilon ^{\prime }t/\hbar} \nonumber  \\
&=&\sum_{m}J_{m}^{2}\left( \alpha _{e,c}\right) \int d\varepsilon ^{\prime
}\Gamma ^{e,c}\left( \varepsilon ^{\prime }\right) \delta \left( m\hbar \Omega
-\varepsilon +\varepsilon ^{\prime }\right) 
\end{eqnarray}%
which indicates that photons of the external a.c. field shift energies of tunneling electrons by $m\hbar \Omega$. Then%
\begin{eqnarray}
w_{+}\left( m\right) -w_{-}\left( m\right) =\int
d\varepsilon J_{m}^{2}\left( \alpha _{e,c}\right) \Gamma ^{e,c}\left(
\varepsilon _{k,e\left( c\right) }^{m}\right) \nonumber \\ 
\left( f\left( \varepsilon
_{e\left( c\right) }^{m}\right) G_{m}^{-}\left( \varepsilon ,\Omega \right) +%
\left[ 1-f\left( \varepsilon _{e\left( c\right) }^{m}\right) \right]
G_{m}^{+}\left( \varepsilon ,\Omega \right) \right) 
\nonumber
\end{eqnarray}
\begin{widetext}
If the energy dependence of $\Gamma^{e,c}\left(\varepsilon _{k,e\left( c\right) }^{m}\right)$ is negligable, one gets%
\begin{eqnarray*}
&&w_{+}^{e,c}\left( n,m \right) -w_{-}^{e,c}\left( n,m \right) 
\\
&=&\Gamma ^{e.c}J_{m}^{2}\left( \alpha _{e,c}\right) \int
d\varepsilon \left( f_{+}^{e\left( c\right) }G^{-}+\left[ 1-f_{-}^{e\left(
c\right) }\right] G^{+}\right)  \\
&=&\Gamma ^{e.c}J_{m}^{2}\left( \alpha _{e,c}\right) \int
d\varepsilon \left( f_{+}^{e\left( c\right) }\left( {\cal K}+i\mathrm{Re}%
{\cal K}\right) +\left( 1-f_{-}^{e\left( c\right) }\right) \left(
{\cal K}-i\mathrm{Re}{\cal K}\right) \right)  \\
&=&\Gamma ^{e.c}J_{m}^{2}\left( \alpha _{e,c}\right) \int
d\varepsilon \left( {\cal K}\left( 1+f_{+}^{e\left( c\right)
}-f_{-}^{e\left( c\right) }\right) +i\mathrm{Re}{\cal K}\left(
f_{+}^{e\left( c\right) }+f_{-}^{e\left( c\right) }-1\right) \right) 
\end{eqnarray*}%
where we used short notations $f\left(\varepsilon _{e\left( c\right)
}^{m\pm }\right) \rightarrow f_{\pm }^{e\left( c\right) }$ and $G^{\pm }={\cal K}\mp i\mathrm{Re}{\cal K}$. Here $\mathcal{K}=\left( G^{-}+G^{+}\right) /2$ is the full electron correlator from which the retarded Green function is obtained as $%
G^{R}(x,t)=-2i\theta (t)\mathrm{Re}\mathcal{K}(x,t)$. The time-averaged partial electric
current $I^{e}$ between the emitter and the quantum well takes the form 
\begin{eqnarray*}
I^{e} &=&e\sum_{n}\zeta _{n}\left( w_{+}^{e}\left(
n,m \right) -w_{-}^{e}\left( n,m \right) \right)  \\
&=&e\Gamma ^{e} \int
d\varepsilon \sum_{nm}J_{m}^{2}\left( \alpha _{e,c}\right)\zeta _{n}[{\cal K}\left( \varepsilon
\right) +{\cal K}\left( \varepsilon \right) \left[ f_{+}^{e\left(
c\right) }-f_{-}^{e\left( c\right) }\right] 
+\mathrm{\mathrm{Im}}{\cal K}\left( \varepsilon \right) \left( 1-f_{+}^{e\left( c\right)
}-f_{-}^{e\left( c\right) }\right) ] \\
&=&e\Gamma ^{e}\sum_{n}\zeta _{n}n+e\Gamma
^{e}  \int d\varepsilon \sum_{nm}J_{m}^{2}\left( \alpha _{e,c}\right)\zeta
_{n}{\cal K}\left( \varepsilon \right) \left[ f_{+}^{e\left(
c\right) }-f_{-}^{e\left( c\right) }\right]  \\
&&+e\Gamma ^{e} \int d\varepsilon
\sum_{nm}J_{m}^{2}\left( \alpha _{e,c}\right) \zeta _{n}\mathrm{Im}{\cal K}\left( \varepsilon
\right) \left( 1-f_{+}^{e\left( c\right) }-f_{-}^{e\left( c\right) }\right) 
\end{eqnarray*}%
where the number of extra electrons in the well is $n=\sum_{k}{\cal K}\left( \varepsilon \right) /\left(2\pi i\right) $. The electron tunneling between the quantum well and the
electrodes causes a time evolution of the probability $\zeta _{n}$
to find $n$ electrons inside the well. The time dependence of $\zeta _{n}\left(
t\right) $ satisfies to the master equation\cite{Averin}%
\begin{equation}
\dot{\zeta}_{n}=w_{n+1}^{-}\zeta _{n+1}+w_{n-1}^{+}\zeta _{n-1}-\left( w_{n}^{+}+w_{n}^{-}\right) \zeta _{n}
\end{equation}
The collector part of the electric current inside the well follows from the equilibrium condition that
\begin{eqnarray*}
I^{e}-I^{c} &=&0=\frac{2}{h}\cdot e\Gamma ^{e} \int d\varepsilon \sum_{n}\zeta _{n}[{\cal K}\left( \varepsilon \right) J_{m}^{2}\left( \alpha _{e}\right) %
\left[ f_{+}^{e}-f_{-}^{e}\right] -\mathrm{Im}{\cal K}\left( \varepsilon
\right) J_{m}^{2}\left( \alpha _{e}\right) \left[ 1-f_{+}^{e}-f_{-}^{e}%
\right] ] \\
&&-e\Gamma ^{c}\int d\varepsilon \sum_{n}\zeta _{n}\left[ \mathrm{%
Im}{\cal K}\left( \varepsilon \right) J_{m}^{2}\left( \alpha _{c}\right) %
\left[ 1-f_{+}^{c}-f_{-}^{c}\right] -{\cal K}\left( \varepsilon \right)
J_{m}^{2}\left( \alpha _{c}\right) \left[ f_{+}^{c}-f_{-}^{c}\right] \right] 
\\
&&-e\left[ \Gamma ^{e} +\Gamma ^{c}\right]
\sum_{n}\zeta _{n}n
\end{eqnarray*}%
The above equation gives
\begin{eqnarray*}
e\left[ \Gamma ^{e} +\Gamma ^{c}\right]
\sum_{n}\zeta _{n}n &=&\frac{2}{h}\cdot \int
d\varepsilon \sum_{n}\zeta _{n}[{\cal K}\left( \varepsilon
\right) \left( e\Gamma ^{e} J_{m}^{2}\left(
\alpha _{e}\right) \left[ f_{+}^{e}-f_{-}^{e}\right] +e\Gamma
^{c}J_{m}^{2}\left( \alpha _{c}\right) \left[ f_{+}^{c}-f_{-}^{c}\right]
\right)  \\
&&-\mathrm{Im}{\cal K}\left( \varepsilon \right) \left( e\Gamma
^{e} J_{m}^{2}\left( \alpha _{e}\right) \left[
1-f_{+}^{e}-f_{-}^{e}\right] +e\Gamma ^{c}J_{m}^{2}\left( \alpha _{c}\right) %
\left[ 1-f_{+}^{c}-f_{-}^{c}\right] \right) ]
\end{eqnarray*}%
The average number of electrons in the well is%
\[
\left\langle n\right\rangle =\sum_{n}\zeta _{n} n + n_{\rm G},
\]%
where $n=\sum_{k}\mathcal{K}\left( \varepsilon \right) /\left(2\pi i\right) $, $\mathcal{K}=\left( G^{-}+G^{+}\right)
/2$ is the full electron correlator, $G^{\pm }$ is the electron Keldysh Green function (\ref{G_pm}). The second term $n_{\rm G}$ is controlled by the gate voltage $V_{\rm G} \neq 0$ applied to the quantum well as shown in Fig.~\ref{fig:Setup_b}. Then one gets the following expression for the average number of electrons in the well%
\begin{eqnarray}
\left\langle n\right\rangle  &=&\frac{2}{h}\frac{1}{\Gamma
^{e} +\Gamma ^{c}}\cdot \int d\varepsilon
\sum_{nm}\zeta _{n}[{\cal K}\left( \varepsilon \right) \left(
\Gamma ^{e} J_{m}^{2}\left( \alpha
_{e}\right) \left[ f_{+}^{e}-f_{-}^{e}\right] +\Gamma ^{c}J_{m}^{2}\left(
\alpha _{c}\right) \left[ f_{+}^{c}-f_{-}^{c}\right] \right)   \nonumber \\
&&+\mathrm{Im}{\cal K}\left( \varepsilon \right) \left( \Gamma ^{e}J_{m}^{2}\left( \alpha _{e}\right) \left[
1-f_{+}^{e}-f_{-}^{e}\right] +\Gamma ^{c}J_{m}^{2}\left( \alpha _{c}\right) %
\left[ 1-f_{+}^{c}-f_{-}^{c}\right] \right) ] + n_{\rm G}  \label{n_f}
\end{eqnarray}%
Equation (\ref{n_f}) actually determines the chemical potential $\mu_{\cal T}
\left( n\right) $ of electrons in the well. In absence of
the SET one gets $f\left( \varepsilon _{+}^{e}\right) =f\left( \varepsilon
_{-}^{e}\right) $ and the 1$^{st}$ term under $\int d\varepsilon $ in Eq.~(\ref{n_f}) vanishes. Then one simply gets
\begin{eqnarray*}
\frac{I^{e}+I^{c}}{2} &=&\frac{e\Gamma ^{e}}{h}\int d\varepsilon \sum_{n}\zeta _{n}%
\left[ {\cal K}\left( \varepsilon \right) J_{m}^{2}\left( \alpha
_{e}\right) \left[ f_{+}^{e}-f_{-}^{e}\right] -\mathrm{Im}{\cal K}\left(
\varepsilon \right) J_{m}^{2}\left( \alpha _{e}\right) \left[
1-f_{+}^{e}-f_{-}^{e}\right] \right]  \\
&&+e\Gamma ^{c}\int d\varepsilon \sum_{n}\zeta _{n}\left[ \mathrm{%
Im}{\cal K}\left( \varepsilon \right) J_{m}^{2}\left( \alpha _{c}\right) %
\left[ 1-f_{+}^{c}-f_{-}^{c}\right] -{\cal K}\left( \varepsilon \right)
J_{m}^{2}\left( \alpha _{c}\right) \left[ f_{+}^{c}-f_{-}^{c}\right] \right] 
\\
&&-e\left[ \Gamma ^{e} -\Gamma ^{c}\right]
\sum_{n}\zeta _{n}n
\end{eqnarray*}%
If one also sets $n_{\rm G}=0$ then
\begin{eqnarray*}
\frac{I^{e}+I^{c}}{2} &=&\frac{e\Gamma ^{e}}{h}\int d\varepsilon \sum_{k,n}\zeta _{n}\left[ {\cal K}\left( \varepsilon \right) J_{m}^{2}\left( \alpha _{e}\right) \left[
f_{+}^{e}-f_{-}^{e}\right] -\mathrm{Im}{\cal K}\left( \varepsilon \right)
J_{m}^{2}\left( \alpha _{e}\right) \left[ 1-f_{+}^{e}-f_{-}^{e}\right] %
\right]  \\
&&+e\Gamma ^{c}\sum_{k,n}\zeta _{n}\left[ \mathrm{Im}{\cal K}\left( \varepsilon \right) J_{m}^{2}\left( \alpha _{c}\right) \left[1-f_{+}^{c}-f_{-}^{c}\right] -{\cal K}\left( \varepsilon \right)
J_{m}^{2}\left( \alpha _{c}\right) \left[ f_{+}^{c}-f_{-}^{c}\right] \right] 
\\
&&-\frac{2e}{h}\frac{\Gamma ^{e} -\Gamma ^{c}%
}{\Gamma ^{e} +\Gamma ^{c}}\cdot \int
d\varepsilon \sum_{nm}\zeta _{n}[{\cal K}\left( \varepsilon
\right) \left( \Gamma ^{e} J_{m}^{2}\left(
\alpha _{e}\right) \left[ f_{+}^{e}-f_{-}^{e}\right] +\Gamma
^{c}J_{m}^{2}\left( \alpha _{c}\right) \left[ f_{+}^{c}-f_{-}^{c}\right]
\right)  \\
&&-\mathrm{Im}{\cal K}\left( \varepsilon \right) \left( e\Gamma
^{e} J_{m}^{2}\left( \alpha _{e}\right) \left[
1-f_{+}^{e}-f_{-}^{e}\right] +e\Gamma ^{c}J_{m}^{2}\left( \alpha _{c}\right) %
\left[ 1-f_{+}^{c}-f_{-}^{c}\right] \right) ]
\end{eqnarray*}%
Using that%
\[
\Gamma _{\mathrm{n}}=\Gamma ^{e\left(c\right) }\left( 1\mp \frac{\Gamma^{e}-\Gamma ^{c}}{\Gamma ^{e}+\Gamma ^{c}}\right) =\frac{2\Gamma ^{e}\Gamma ^{c}}{\Gamma ^{e}+\Gamma ^{c}}
\]%
the d.c. photon-assisted single electron tunneling electric current (PASET) across the quantum well reads%
\begin{eqnarray}
I &=&\frac{I^{e}+I^{c}}{2}=\Gamma _{\mathrm{n}} \frac{2e}{h}\int d\varepsilon \sum_{nm}\zeta
_{n}[{\cal K}\left( \varepsilon \right) \left[ J_{m}^{2}\left(
\alpha _{e}\right) \left( f_{+}^{e}-f_{-}^{e}\right) -J_{m}^{2}\left( \alpha
_{c}\right) \left( f_{+}^{c}-f_{-}^{c}\right) \right]   \label{net_curr} \\
&&+\mathrm{Im}{\cal K}\left( \varepsilon \right) \left[ J_{m}^{2}\left(
\alpha _{e}\right) \left( f_{+}^{e}+f_{-}^{e}\right) -J_{m}^{2}\left( \alpha
_{c}\right) \left( f_{+}^{c}+f_{-}^{c}\right) \right] ]  \nonumber
\end{eqnarray}%
In equilibrium one makes use the Fourier transform
\[
{\cal K}_0 \left({\bf k}\right)=\frac{i}{2}\tanh \left( \frac{\beta v_{F}|{\bf k}| }{2}\right) =%
\frac{i}{2}\left( 1-2{\cal G}_0(\beta v_{F}|{\bf k}|)\right) 
\]
of the non-interactive equilibrium right-moving fermion correlator ${\cal K}_{0}\left(
0,t\right) $%
\[
{\cal K}_{0}\left( x,t\right) =\frac{i}{x-v_{F}t+i\delta }\frac{\pi \left(
x-v_{F}t\right) /\beta v_{F}}{\sinh \left( \pi \left( x-v_{F}t\right) /\beta
v_{F}\right) }
\]
where ${\cal G}_0(\varepsilon )=1/(\exp{(\varepsilon/T)}+1)$, $\bf k$ is the fermion momentum,  $\delta \rightarrow +0$. A conventional way\cite{Elesin} to introduce the electron distribution function ${\cal G}\left(\varepsilon \right) $ is to apply the anzatz
\begin{equation}
{\cal K}\left( \varepsilon \right) =\mathrm{Im}{\cal K}\left(
\varepsilon \right) \left( 1-2{\cal G}\left( \varepsilon \right) \right) 
\label{G_f}
\end{equation} 
for the net time-averaged electric current across the double-barrier junction one gets%
\begin{equation}
I=\Gamma _{\mathrm{n}} \frac{2e}{h}\int
d\varepsilon \sum_{nm}\zeta _{n}\mathrm{Im}{\cal K}\left(
\varepsilon \right) [J_{m}^{2}\left( \alpha _{e}\right) \left[
f_{+}^{e}+{\cal G}\left( f_{-}^{e}-f_{+}^{e}\right) \right]
-J_{m}^{2}\left( \alpha _{c}\right) \left[ f_{+}^{c}+\left(
f_{-}^{c}-f_{+}^{c}\right) {\cal G}\right] ].
\label{curr_SET}
\end{equation}
From Eq.~(\ref{curr_SET}) in limits $\alpha_{e,c}=0$ or ${\cal G}(\varepsilon )=0$ one easily recovers the expressions for tunneling current used in the main text. 
\end{widetext}


\begin{thebibliography}{99}
\bibitem{Dressel} M. S. Dresselhaus, G. Dresselhaus, and P. Avouris, Carbon
nanotubes: synthesis, structure, properties, and applications, (Springer,
New York, 2001).

\bibitem{chem} E. S. Snow, F. K. Perkins, E. J. Houser, S. C. Badescu, T. L.
Reinecke, Science \textbf{307},1942 (2005).

\bibitem{med} V. P. Wallace, A. J. Fitzgerald, S. Shankar, R. J. Pye, D. D.
Arnone, Brit. J. Derm. \textbf{151}, 424 (2004).

\bibitem{My-PRB} S. E. Shafranjuk, Phys. Rev. B\textbf{76}, 085317 (2007).

\bibitem{Paola-A} J. Zhang, A. Boyd, A. Tselev, M. Paranjape, and P.
Barbara, Applied Physics Letters \textbf{88}, 123112 (2006); A. Tselev, K.
Hatton, M. S. Fuhrer, M. Paranjape, and P. Barbara, Nanotechnology \textbf{15%
}, 1475 (2004).

\bibitem{Paola-B} J. Zhang, A. Tselev, Y. Yang, K. Hatton, P. Barbara and S.
Shafraniuk, Phys. Rev. B \textbf{74}, 155414 (2006).

\bibitem{Lambert} D. Gunlycke, C. J. Lambert, S. W. D. Bailey, D. G.
Pettifor, G. A. D. Briggs, and J. H. Jefferson, Europhys. Lett. \textbf{73},
759 (2006).

\bibitem{Babic} B. Babic and C. Sch\"{o}nenberger, Phys. Rev. B \textbf{70},
195408 (2004

\bibitem{Cao} J. Cao, Q. Wang, H. Dai, Nature Materials \textbf{4}, 745
(2005).

\bibitem{Wiel} W. G. van der Wiel, S. De Franceschi, T. Fujisawa, J. M.
Elzerman, S. Tarucha, and L. P. Kouwenhoven, Science \textbf{289}, 2105
(2000).

\bibitem{Mintmire} J.W. Mintmire and C. T. White, Phys. Rev.
Lett. \textbf{81}, 2506 (1998).

\bibitem{Kane-PRL} C. L. Kane, L. Balents and M.P.A. Fisher, Phys. Rev.
Lett. \textbf{79}, 5086 (1997).

\bibitem{Egger} R. Egger and A. O. Gogolin, , Phys. Rev. Lett. \textbf{79}, 5082 (1997).

\bibitem{TLL} S. Tomonaga, Prog. Theor. Phys. \textbf{5}, 544 (1950); J.M.
Luttinger, J. Math. Phys. (N.Y.) \textbf{4}, 1154 ( 1963).

\bibitem{Bockrath} M. Bockrath et al., Nature (London) \textbf{397}, 598
(1999); Yao et al., Nature (London) \textbf{402}, 273 (1999).

\bibitem{Ishii} H. Ishii, H. Kataura, H. Shiozawa, H. Yoshioka, H. Otsubo,
Y. Takayama, et al. Letters to Nature, \textbf{426}, 540 (2003).

\bibitem{LL-recent} P. Recher, N. Y. Kim, and Y. Yamamoto, Phys. Rev. B 
\textbf{74}, 235438 (2006).

\bibitem{Nazarov} Yu. V. Nazarov and L. I. Glazman, Phys. Rev. Lett. \textbf{%
91}, 126804 (2003).

\bibitem{Sonin} R. Tarkiainen, M. Ahlskog, J. Penttila, L. Roschier, P.
Hakonen, M. Paalanen, and E. Sonin, Phys. Rev. B \textbf{64}, 195412 (2001).

\bibitem{Emery} See V. Emery, in Highly Conducting One-Dimensional Solids,
edited by J. Devreese et al. (Plenum, New York, 1979).

\bibitem{Keldysh} L. V. Keldysh, Sov. Phys. JETP \textbf{20}, 1018 (1965).

\bibitem{Datta} S. Datta, Electronic Transport in Mesoscopic Systems
(Cambridge University Press, Cambridge, UK, 1997).

\bibitem{Averin} D. V. Averin, A. N. Korotkov, and K. K. Likharev, Phys.
Rev. B \textbf{44}, 6199 (1991).

\bibitem{Feldman} D. E. Feldman, S. Scheidl, and V. M. Vinokur, Phys. Rev.
Lett. \textbf{94}, 186809 (2005).

\bibitem{Trushin} M. Trushin, A. L. Chudnovskiy, Europhys. Lett. \textbf{82}%
, 17008 (2008).

\bibitem{Elesin}V. F. Elesin and Yu. V. Kopaev, Sov. Phys. Uspekhi, {\bf  24}, 116 (1981).

\bibitem{Keeffe} J. O$^\prime $Keeffe, C. Wei, and K. Cho, Appl. Phys.
Lett., \textbf{80}, 676 (2002).

\bibitem{Chen} C.-W. Chen, M.-H. Lee, and S. J. Clark, Nanotechnology, 
\textbf{15}, 1837 (2004).

\bibitem{Chiu} P. W. Chiu, M. Kaempgen, and S. Roth, Phys. Rev. Lett., 
\textbf{92}, 246802 (2004).

\bibitem{Li} Y. Li, S. V. Rotkin, and U. Ravaioli, Nanoletters, \textbf{3},
183 (2003).
\end{thebibliography}
\end{document}